\documentstyle[preprint,prd,aps]{revtex}
\newcommand{\be}{\begin{equation}}
\newcommand{\ee}{\end{equation}}
\newcommand{\bea}{\begin{eqnarray}}
\newcommand{\eea}{\end{eqnarray}}
\begin{document}
\draft
\title{Charged spin $\frac{1}{2}$ particle in an arbitrary magnetic field in two spatial dimensions: \\
a supersymmetric quantum mechanical system}
\author{T.E. Clark\footnote{e-mail address: clark@physics.purdue.edu}, S.T. Love\footnote{e-mail address: love@physics.purdue.edu} and S.R. Nowling\footnote{e-mail address: snowling@physics.purdue.edu}}
\address{\it Department of Physics, 
Purdue University,
West Lafayette, IN 47907-1396}
\maketitle
\begin{abstract}
It is shown that the $2\times 2$ matrix Hamiltonian describing the dynamics of a charged spin $1/2$ particle with $g$-factor $2$ moving in an arbitrary, spatially dependent, magnetic field in two spatial dimensions can be written as the anticommuator of a nilpotent operator and its hermitian conjugate. Consequently, the Hamiltonians for the two different spin projections form partners of a supersymmetric quantum mechanical system. The resulting supersymmetry algebra can then be exploited to explicitly construct the exact zero energy ground state wavefunction for the system. Modulo this ground state, the remainder of the eigenstates and eigenvalues of the two partner Hamiltonians form positive energy degenerate pairs. We also construct the spatially asymptotic form of the magnetic field which produces a finite magnetic flux and associated zero energy normalizable ground state wavefunction. 
\end{abstract}

\newpage

The calculation of the stationary states of an electron moving in an uniform magnetic field has been well studied for over 70 years and by now constitutes a classic textbook example\cite{LL}. The consequent Landau level energies and associated wavefunctions characterizing the dynamics in the two dimensional plane normal to the magnetic field have important applications in condensed matter systems ranging from the de Hass-van Alphen effect in metals\cite{condmatt} to the quantum Hall effect\cite{qhe}. Moreover, it has been observed\cite{J}-\cite{K} that when the particle has Land\'e factor $g=2$, the system exhibits a supersymmetry with the Hamiltonians corresponding to the spin up and spin down projections forming supersymmetric partners. The resulting supersymmetry algebra can be exploited to compute the exact zero energy ground state wavefunction as well as to show that the positive energy eigenvalues of the two Hamiltonians form degenerate pairs. It has also been previously noted\cite{R} that the model describing the dynamics of a charged spin 1/2 particle with Land\'e factor $g=2$ continues to exhibit a supersymmetry even for arbitrary, spatially dependent, magnetic fields in two spatial dimensions. In the present note, we elaborate upon this observation and address some of its consequences. After explicitly constructing the supersymmetry charges and demonstrating the supersymmetry algebra, we use the construct to explicitly secure the exact form of the ground state zero energy wavefunction. The energy eigenvalues of the two supersymmetric partner Hamiltonians, which again correspond to the spin up and spin down projections, form degenerate pairs, save for the zero energy ground state of only one of the Hamiltonians. Finally, we discuss necessary conditions to produce normalizable ground states and finite magnetic field flux out of the plane.

The Hamiltonian for a spin $\frac{1}{2}$ particle of charge $q$ and mass $m$ moving in the $x-y$ plane under the influence of an arbitrary, spatially dependent, static magnetic field in the $z$-direction, $\vec{B} =B(x,y) \hat{z}$, is given by
\be
H=\frac{1}{2m}\left( \vec{p} -\frac{q}{c}\vec{A}(x,y)\right)^2 - g \frac{q}{2mc}B(x,y)S ~,
\ee
with the spin operator $S=\frac{\hbar}{2} \sigma_3 =\frac{\hbar}{2} \pmatrix{1&0\cr
0&-1}$ and the spin magnetic moment characterized by the Land\'e $g$-factor.  In two dimensions, the most general form for the magnetic vector potential $\vec{A}$ is ($i,j = 1,2$)
\be
A_i(x,y) = \partial_i C(x,y) + \epsilon_{ij}\partial_j K(x,y)~, 
\ee
while the magnetic field is given by the curl of $\vec{A}$ as
\be
B(x,y) = \epsilon_{ij} \partial_i A_j (x,y) = - \nabla^2 K(x,y)~.
\ee
Adding a gradient to the magnetic vector potential leaves the magnetic field unaltered and corresponds to a gauge transformation. This implies that the scalar function $C(x,y)$ is arbitrary while the prepotential $K(x,y)$ is defined up to the addition of harmonic functions. Thus $K(x,y)$ and $K^\prime (x,y) = K(x,y) +F(x+iy) +F^*(x-iy)$ produce identical magnetic fields since the holomorphic functions $F(x+iy)$ and $F^*(x-iy)$ automatically satisfy $\nabla^2 F(x+iy) = 0 = \nabla^2 F^*(x-iy)$. As usual, all physical quantities such as the energies are necessarily gauge invariant. In the following, we choose to work in the Coulomb gauge defined by $\vec{\nabla} \cdot \vec{A} =0$ and with $C=0$.  The magnetic vector potential is then purely solenoidal and of the form $A_i (x,y) = \epsilon_{ij} \partial_j K(x,y)$, where $K$ still retains the residual gauge freedom of the addition of $F(x+iy) +F^*(x-iy)$.

It proves convenient to introduce complex coodinates $x_\pm = x\pm iy$, 
momenta $p_\pm = \frac{1}{2}\left(p_x \mp ip_y \right) =\frac{\hbar}{i}\frac{\partial}{\partial x_\pm}\equiv \frac{\hbar}{i}\partial_\pm $
and magnetic vector potential components $A_\pm = \frac{1}{2}\left( A_x \pm i A_y \right) = \mp i \partial_\mp K$, where the  prepotential is now considered a function of $x_\pm$ so that $K= K(x_+,x_-)$.  It follows that the magnetic field then can be written as $B(x,y) = -4 \partial_{+}\partial_{-}K(x_+,x_-)$, while the momenta conjugate to $x_\pm$ take the form
\be
\pi_\mp = 2\left( p_\mp \pm i\frac{q}{c} \partial_\mp K\right) = (\pi_\pm)^\dagger
\ee
and have the commutator $\left[ \pi_-, \pi_+ \right] = -8\frac{\hbar q}{c}\partial_+\partial_- K(x_+,x_-) =2\frac{\hbar q}{c}B(x,y)$.

When the Land\'e $g$-factor takes the value $g=2$, the Hamiltonian factorizes as 
\be
H = \pmatrix{ H_\uparrow &0 \cr
0& H_\downarrow} =\frac{1}{2m}\pmatrix{\pi_+ \pi_- & 0 \cr
0& \pi_- \pi_+ }.
\ee
Here $H_\uparrow = \frac{\pi_+ \pi_-}{2m} (H_\downarrow = \frac{\pi_- \pi_+}{2m}) $ are the respective Hamiltonians for the spin projections $+\frac{\hbar}{2} (-\frac{\hbar}{2})$. This matrix Hamiltonian can further be written as the square of a hermitian operator $\cal Q$ as
\be
H = \frac{1}{2} {\cal Q}^2 ~;~  {\cal Q} = \frac{1}{\sqrt{m}}\pmatrix{ 0& -i \pi_+ \cr
i\pi_- & 0} = ({\cal Q})^\dagger,
\ee
An immediate consequence of this observation is that the energy spectrum is necessarily non-negative.

Using the Pauli matrices, $\sigma_+ = \frac{1}{2}(\sigma_1+i\sigma_2)=\pmatrix{0&1\cr
0&0}~~;~~
\sigma_- =\frac{1}{2}(\sigma_1-i\sigma_2)= \pmatrix{0&0\cr
1&0}$, the operator ${\cal Q}$ can be further written as the sum ${\cal Q} = Q + {Q}^\dagger$, where 
\be
Q = \frac{i}{\sqrt{m}} \pi_- \sigma_- ~;~
Q^\dagger = \frac{-i}{\sqrt{m}} \pi_+ \sigma_+ 
\ee
are two complex, nilpotent, $Q^2=0=(Q^\dagger)^2$, supersymmetry charges. 
These charges, together with the Hamiltonian, obey the supersymmetry algebra \cite{Witten}\cite{Rosner}\cite{Khare}
\bea
\left\{ Q, Q\right\} = &0&=\left\{ {Q}^\dagger, {Q}^\dagger\right\}\cr
\left\{ Q, {Q}^\dagger\right\} &=& 2H \cr
\left[ Q, H \right] = &0& = \left[{Q}^\dagger, H\right] .
\eea

The general normalizable $H_{\uparrow}$ eigenstate, $\psi_{\uparrow n}$, has energy $E_{\uparrow n} \ge 0$ and satisfies 
\be
H_{\uparrow}\psi_{\uparrow n}=\frac{\pi_+ \pi_-}{2m} \psi_{\uparrow n}=E_{\uparrow n}\psi_{\uparrow n}~~;~~E_{\uparrow n}\ge 0 .
\ee
Suppose $H_{\uparrow}$ has the normalizable zero energy ($E_{\uparrow 0}=0$) eigenstate
$\psi_{\uparrow 0}$ satisfying $H_{\uparrow}\psi_{\uparrow 0}=0$ which implies that $\pi_-\psi_{\uparrow 0}=0 $. The vanishing commutator of $Q$ and $Q^\dagger$ with $H$ implies that $\pi_- H_\uparrow = H_\downarrow \pi_- $ and $\pi_+ H_{\downarrow}=H_\uparrow \pi_+$. It follows that left multiplication of the $H_{\uparrow}$ eigenvalue equation by $\pi_-$ then dictates that $H_{\downarrow }(\pi_- \psi_{\uparrow n})=E_{\uparrow n} (\pi_- \psi_{\uparrow n})$, so that $\pi_- \psi_{\uparrow n}~~,~~ n>0,$ is an $H_{\downarrow}$ eigenstate with eigenvalue $E_{\uparrow n} > 0$. Note that the case $n=0$ does not give an $H_{\downarrow}$ eigenstate since $\pi_- \psi_{\uparrow 0}=0$. Thus the normalizable ground state of $H_{\downarrow}$ is $\psi_{\downarrow 0}=N_{\downarrow 0}\pi_- \psi_{\uparrow 1}$, where $N_{\downarrow 0}$ is a normalization constant, and has energy $E_{\uparrow 1}$. Except for the zero energy eigenstate of $H_{\uparrow}$, all the other eigenstates of $H_{\uparrow}$ and $H_{\downarrow}$ pair up with the same positive energy eigenvalues. This is a direct consequence of the supersymmetry. Thus if $\psi_{\uparrow n}$ is an eigenstate of $H_{\uparrow}$ with eigenvalue $E_{\uparrow n}$, then $\psi_{\downarrow n} = N_{\downarrow n} \pi_- \psi_{\uparrow n+1}~~,~~ n \ge 0,$ is an eigenstate of $H_{\downarrow}$ with eigenvalue $E_{\downarrow n}=E_{\uparrow n+1}>0$. 

To explicitly construct the zero energy ground state of $H_\uparrow$, we use that 
\be
\pi_- \psi_{\uparrow 0}= 2(\frac{\hbar}{i}\partial_-  +i\frac{q}{c}\partial_- K)\psi_{\uparrow 0}=0 ~.
\ee
This differential equation is readily solved yielding
\be
\psi_{\uparrow 0}(x_+,x_-)=N_{\uparrow 0}U(x_+)e^{\frac{q}{\hbar c}K(x_+,x_-)}~.
\ee
Here $U(x_+)$ is an arbitrary complex function which reflects the degeneracy of the state. This zero energy eigenstate will be the system ground state provided it is normalizable.

On the other hand, suppose the $H_{\downarrow}$ eigenvalue equation
\be
H_{\downarrow}\psi_{\downarrow n}=E_{\downarrow n}\psi_{\downarrow n}~~;~~
E_{\downarrow n} \ge 0
\ee
admits the normalizable zero energy eigenstate, $\psi_{\downarrow 0}$, satisfying $\pi_+ \psi_{\downarrow 0} =0$ and $E_{\downarrow 0}=0$. Then an analogous argument gives that the normalizable $H_{\uparrow}$ ground state as $\psi_{\uparrow 0}=N_{\uparrow 0} \pi_+ \psi_{\downarrow 1}$ having positive energy $E_{\uparrow 0}=E_{\downarrow 1}$. Once again, except for the zero energy eigenstate of $H_{\downarrow}$, all the other normalizable eigenstates of $H_{\downarrow}$ and $H_{\uparrow}$ pair up with degenerate positive energy eigenvalues. Thus  $\psi_{\uparrow n}=N_{\uparrow n}\pi_+ \psi_{\downarrow n+1}$ is an $H_{\uparrow}$ eigenstate with eigenvalue $E_{\uparrow n}=E_{\downarrow n+1}>0$. In this case, the zero energy ground state is gleaned from the condition 
\be
\pi_+ \psi_{\downarrow 0}=2(\frac{\hbar}{i}\partial_+-\frac{iq}{c}\partial_+ K)\psi_{\downarrow 0}=0
\ee
whose solution is
\be
\psi_{\downarrow 0}(x_+,x_-)=N_{\downarrow 0}U(x_-)e^{-\frac{q}{\hbar c}K(x_+,x_-)}
\ee
with $N_{\downarrow 0}$ a normalization constant. This zero energy eigenstate will be the system ground state provided it is normalizable. If $K$ is such that neither of the zero energy wavefunctions $U(x_+)e^{\frac{q}{\hbar c}K(x_+,x_-)}$ nor $U(x_-)e^{-\frac{q}{\hbar c}K(x_+,x_-)}$ are normalizable, then all states, including the ground states, have paired positive energies. This corresponds to a case of broken supersymmetry.

These general results, obtained for an arbitrary $x,~y$ dependent magnetic field may also be applied to the example of a uniform magnetic field: $B > 0$.  A corresponding  prepotential is given by
\be
K(x_+,x_-)=-\frac{1}{2}B(\frac{x_+-x_-}{2i})^2=-\frac{1}{2}B y^2 .
\ee
Note that this choice of $K$ corresponds to the asymmetric gauge choice (still in Coulonb gauge) $A_x =  -B y ~;~ A_y =  0$. Recalling that $K$ is only fixed up to additive functions $F(x_+) + F(x_-)$, an alternate choice for $K$ is $K(x_+.x_-)=-\frac{1}{4}Bx_+x_- = -\frac{1}{4}B(x^2+y^2)$ which corresponds to the symmetric gauge 
$A_x= -\frac{1}{2}By~;~A_y= \frac{1}{2}Bx$.  This is related to the asymmetric gauge choice by the holomorphic gauge transformation functions $F(x_+)=\frac{1}{8}Bx_+^2$.

Working in the asymmetric (Landau) gauge, the state 
\be
\psi_{\uparrow 0}(x_+,x_-)=N_{\uparrow 0}U(x_+) e^{-\frac{qB}{2\hbar c}(\frac{x_+-x_-}{2i})^2}=N_{\uparrow 0}U(x+iy)e^{-\frac{qB}{2\hbar c}y^2}
\ee
is the normalizable zero energy ground state. This is the familiar lowest Landau level eigenstate. Since any function can be expanded in terms of plane waves, we can choose 
$U(x_+)=e^{ikx_+}$ and write
\be
\psi_{\uparrow 0 k}(x,y) =N_{\uparrow 0 k} e^{ik(x+iy)}e^{-\frac{qB}{2\hbar c}y^2}
\ee
where we have included a label $k$ on the wavefunction which labels the degeneracy. Note that this wavefunction can be rewritten after completing the square as $\psi_{\uparrow 0 k}(x,y)=N_{\uparrow 0 k} e^{ikx}e^{-\frac{qB}{2\hbar c}(y-y_0)^2}$ where $y_0=-\frac{\hbar c k}{qB}$ and the overall normalization constant has been changed. For an unbounded system, the wavefunction, in this gauge, only exhibits a continuum normalization in the $x$-direction, while the level is infinitely degenerate as there is no constraint on the allowed $k$ and $y_0$ values. If the system is of finite extent, say a box of area $L^2$, with $L>>\sqrt{\frac{\hbar c}{qB}}$, and periodic boundary conditions are imposed in the $x$ direction so that $\psi_{\uparrow 0 k}(x=-L/2, y)=\psi_{\uparrow 0 k}(x=L/2, y)$, then the number of degenerate states is $\frac{L}{2\pi}\Delta k=\frac{L}{2\pi}\frac{qB}{\hbar c}\Delta y_0=\frac{L^2}{2\pi}\frac{qB}{\hbar c}$ where we have used that $\Delta y_0 = L$. Note that no particular boundary condition has been imposed on the $y$ direction. We are tacitly assuming vanishing boundary conditions and although the wavefunction really does not vanish at $y=\pm \frac{L}{2}$, it is exponentially small there. On the other hand, the state $e^{-ikx_-}e^{\frac{qB}{2\hbar c}y^2}$ is clearly non-normalizable (in the continuum) so $H_{\downarrow}$ does not admit a zero energy normalizable eigenstate (all its eigenstates have positive energy). Note further that if the sign of $B$ is reversed or if the sign of $q$ is reversed (but not both) the role of $H_{\uparrow}$ and $H_{\downarrow}$ are reversed and it is then $H_{\downarrow}$ which has the normalizable zero energy eigenvalue. 

In this example, one can, in fact, extract the entire spectrum. Noting that $[\frac{\pi_-}{\sqrt{2m}}, \frac{\pi_+}{\sqrt{2m}}]=\hbar \omega_c$, with $\omega_c = \frac{q B}{mc}$, is precisely the commutation relation for the raising and lowering operators of a one dimensional simple harmonic oscillator, it follows that $[H_{\uparrow}, \pi_+]=\hbar \omega_c \pi_+$ 
and consequently $E_{\uparrow n}=n\hbar \omega_c  ~;~ n=0,1,2,...$ (note absence of zero point energy), while $\psi_{\uparrow~n}=N_{\uparrow~n}(\pi_+)^n \psi_{\uparrow~0}~;~n=0,1,2,...$ Moreover, since $H_{\uparrow}$ and $H_{\downarrow}$ only differ by a constant, $H_{\uparrow}- H_{\downarrow} = -\hbar \omega_c $, then $E_{\downarrow~n}=E_{\uparrow~n}+ \hbar \omega_c = E_{\uparrow~n+1}=(n+1)\hbar \omega_c ~;~n=0,1,2,...$ and $\psi_{\downarrow~n}=\psi_{\uparrow~n+1}$. 

Note that in this uniform $B$ field case, the spectrum can also be gleaned even in the case when $g \ne 2$ and the supersymmetry is broken. For any value of $g$, the Hamiltonians $H_{\uparrow}$ and $H_{\downarrow}$ are simply modified by the additional constants $-(g-2)\hbar \omega_c$ and $(g-2)\hbar\omega_c$. The resulting energy spectra for the spin up and spin down projections (assuming the spin up projection has the zero energy ground state when $g=2$) are $E_{\uparrow n} = \hbar\omega_c[n-(g-2)]$ and $E_{\downarrow n} = \hbar \omega_c[n+1+(g-2)]$. Note that for $g$ values other than 2, the ground state ($n=0$) energy no longer vanishes relecting the broken supersymmetry. For the special value, $g=3/2$, the two Hamiltonians are identical and their energy eigenvalues are $\hbar \omega_c (n+\frac{1}{2})$, while for $g=1$, the roles of $H_{\uparrow}$ and $H_{\downarrow}$ are reversed. Moreover, if $g<1$ or $g>2$, the spectrum contains a finite number of negative energy states.

Let us return to the case of an arbitrary magnetic field and unbounded two dimensional space. In order for the zero energy eigenstate to be normalizable, it is necessary that the prepotential magnitude diverges at large distances. On the other hand, in order for the $B$ field flux out of the $x-y$ plane, 
\be
\Phi = \int_0^\infty d\rho ~\rho \int_0^{2\pi} d\varphi ~B =\int_0^{2\pi}d\varphi [\rho A_\varphi]|_{\rho\rightarrow \infty}
\ee  
to be finite, the $B$ field is required to fall off at large distances faster than  $\frac{1}{\rho^2}=\frac{1}{x^2+y^2}$. Here we have introduced the plane polar coordinates, $\rho, \varphi$, where $x=\rho cos\varphi,~ y=\rho sin\varphi$, so that $x_\pm = \rho e^{\pm i\varphi}$, $A_\varphi$ is the $\varphi $ component of $\vec{A}$ and have employed Stokes' theorem to obtain the last equality. Taken together, in order to have both a normalizable zero energy ground state wavefunction and a finite $B$ field flux, the prepotential must be proportional to powers of logarithms. A more detailed analysis reveals that the asymptotic behavior for $K(\rho,\varphi)$ which meets both of the stated criteria can be written\cite{J} as
\be
K(\rho, \varphi) \sim -\frac{q_M}{8}\ell n(\frac{x_+ x_-}{\rho_0^{2}}) = -\frac{q_M}{4}\ell n (\frac{\rho}{\rho_0})~;~q_M > 0 . 
\ee
Here $\rho_0$ is an arbitrary length scale in the logarithm. Since any value of $\rho_0$ is as good as any other and thus no physical observable can depend on the $\rho_0$, we can simply take $\rho_0 = \frac{q_M^2}{\hbar}$ since it carries dimension length. The normalizable zero energy ground state wavefunctions are
\be
\psi_{\uparrow 0}(\rho, \varphi)_n\sim N_{\uparrow 0~n}\rho^n e^{in\varphi}e^{-\frac{q q_M}{4\hbar c}\ell n (\frac{\hbar \rho}{q_M^2})}= N_{\uparrow 0~n}(\frac{\hbar}{q_M^2})^{\frac{q q_M}{4\hbar c}}e^{in\varphi}
\rho^{n-\frac{q q_M}{4\hbar c}}, 
\ee
where we have taken $U(x+iy)=(x+iy)^n =\rho^n e^{in\varphi}$, with $n$ a non-negative integer labeling the degeneracy. In fact, normalizability requires the parameters to satisfy $n<\frac{qq_M}{4\hbar c}-1$. Thus the ground state degeneracy, $n+1$, is given by the largest integer less than $\frac{qq_M}{4\hbar c}$. 

Prepotentials depending on $\ell n{\rho}$ to a higher (lower) power give divergent (vanishing) $B$ flux while producing normalizable (non-normalizable) zero energy wavefunctions. The above prepotential, in turn, corresponds to the asymptotic vector potential components:
\bea
A_\rho &= &A_x cos \varphi +A_y sin \varphi \sim 0     \cr
A_\varphi &=& -A_x sin \varphi + A_y cos \varphi \sim \frac{q_M}{4\rho}~.
\eea
The resulting magnetic field strength then vanishes asymptotically
\be
B^{(1)}(\rho, \varphi)\sim 0
\ee
while producing the magnetic flux
\be
\Phi = \frac{\pi}{2}q_M .
\ee

A physical configuration exhibiting the above asymptotic vector potential is that of a vortex\cite{V}. For a vortex characterized by the (quantized) magnetic flux $N\frac{2\pi \hbar c}{q}~,~N=1,2,...$, it follows that $\frac{qq_M}{4\hbar c}=N$ and the ground state degeneracy is simply given by $N-1$. Note that this implies that the  vortex solution carrying a single unit of magnetic flux, $N=1$, does not produce a normalizable zero energy ground state in the unbounded plane. This is a straightforward consequence of the fact that the zero energy wavefunction in the presence of such a vortex falls asymptotically only as $1/\rho$. An analogous result appears when the charged, spin $1/2$ particle is described by a relativistic massless Dirac equation\cite{J}. In this case, the underlying connection between the integer characterizing the quantized magnetic flux of the vortex and the zero energy ground state degeneracy has a natural interpretation in terms of an index theorem denumerating the number of zero modes of the Dirac differential operator in a magnetic vortex background. Here we have explicitly seen that the zero energy ground state degeneracy is again related to the number of integer units of magnetic flux when the spin $1/2$ particle is described by a non-relativistic Schr\"odinger equation where there is no index theorem connecting the two integer values.

\bigskip

\noindent
This work was supported in part by the U.S. Department 
of Energy under grant DE-FG02-91ER40681 (Task B).

\end{document}